\documentclass[twocolumn,prb,aps,superscriptaddress,showpacs]{revtex4}%
\usepackage{amsmath}
\usepackage{amssymb}
\usepackage{amsfonts}
\usepackage{bm}
\usepackage{amssymb}
\usepackage[dvips]{graphicx}
\usepackage{dcolumn}
\usepackage{txfonts}
\usepackage{hyperref}
\usepackage{makeidx}
\usepackage{color}
\usepackage{mathtools}

\begin{document}

\title{Dynamical decoupling design for identifying weakly coupled nuclear spins in a bath}
\author{Nan Zhao}
\affiliation{Department of Physics and Centre for Quantum Coherence, The Chinese University of Hong Kong, Shatin,
New Territories, Hong Kong, China}
\affiliation{3. Physikalisches Institut and Research Center SCOPE, University of Stuttgart, Pfaffenwaldring 57, 70569 Stuttgart, Germany}
\affiliation{Beijing Computational Science Research Center, Beijing 100084, China}

\author{J\"{o}rg Wrachtrup}
\affiliation{3. Physikalisches Institut and Research Center SCOPE, University of Stuttgart, Pfaffenwaldring 57, 70569 Stuttgart, Germany}

\author{Ren-Bao Liu}
\affiliation{Department of Physics and Centre for Quantum Coherence, The Chinese University of Hong Kong, Shatin,
New Territories, Hong Kong, China}


\begin{abstract}
Identifying weakly coupled nuclear spins around single electron spins is a key step of implementing quantum information processing using coupled electron-nuclei spin systems 
or sensing like single spin nuclear magnetic resonance detection using diamond defect spins.
Dynamical decoupling control of the center electron spin with periodic pulse sequences [e.g., the Carre-Purcell-Meiboom-Gill (CPMG) sequence]
has been successfully used to identify single nuclear spins and to resolve structure of nuclear spin clusters. 
Here, we design a new type of pulse sequences by replacing the repetition unit (a single $\pi$-pulse) of the CPMG sequence with a group of nonuniformly-spaced $\pi$-pulses.
Using nitrogen-vacancy center system in diamond, we show that the designed pulse sequence improves the resolution of nuclear spin noise spectroscopy, 
and more information about the surrounding nuclear spins is extracted.
The principle of dynamical decoupling design proposed in this paper is useful in many systems (e.g., defect spin qubit in solids, trapped ion and superconducting qubit) for high-resolution noise spectroscopy.
\end{abstract}

\maketitle

\section{Introduction}
Coupled electron-nuclear spin systems are important platform for quantum information processing\cite{Childress2006,Dutt2007,Neumann2008,Jiang2009,Neumann2010}.
Single electron spins are promising quantum processors because of their addressability and controllability\cite{Jelezko2004,Jelezko2004b}. 
Nuclear spins are regarded as ideal quantum memories since they are less sensitive to the environmental noise and have longer coherence time\cite{Maurer2012a}. 
Coupled electron-nuclear spin systems have the advantages of both ingredients.
However, in many solid state systems, a large number of nuclear spins around the electron spin usually form a spin bath, 
serving as a decoherence source of electron spins, rather than a kind of resource.
Individual nuclear spins have to be resolved, otherwise they can hardly be used as quantum memories.
In this sense, identifying single nuclear spins in a spin bath is highly desirable.

Single nuclear spins can be resolved through the splitting of the electron spin resonance (ESR) spectrum due to coupling to the nuclear spins.
The resolution of the ESR spectrum splitting is limited by the transition linewidth.
Usually, only the nuclear spins strongly coupled to the electron spin (with the coupling strengths much larger than the linewidth) can be well resolved\cite{Childress2006,Dutt2007,Neumann2008,Jiang2009,Neumann2010}.
Resolving the weakly coupled nuclear spins (with coupling strengths comparable to the linewidth) is a challenging task. 
This greatly limits the potential of using these nuclear spins as resources.

The limitation on resolution by ESR linewidth can be overcome by actively controlling the electron spins via, for example, 
continuously driving spectroscopy\cite{Cai2013,Cai2011a, London2013} and dynamical decoupling (DD) control\cite{Zhao2011a, Zhao2012}.
In particular, DD control of electron spins was proposed to be a powerful tool to detect small nuclear spin clusters\cite{Zhao2011a}.
Under DD control, single nuclear spin clusters around center electron spins manifest themselves as characteristic fingerprint oscillations on the electron spin coherence.
With this discovery, nuclear magnetic resonance can be realized at the single-molecule level\cite{Zhao2011a}. 
Very recently, applying the standard Carre-Purcell-Meiboom-Gill (CPMG) DD sequence on the electron spins of nitrogen-vacancy (NV) centers, 
people have successfully demonstrated the identification of weakly coupled individual nuclear spins\cite{Kolkowitz2012a,Taminiau2012a,Zhao2012}, and the resolving of the structure of single nuclear spins cluster\cite{Shi2013}.


In addition to the widely used periodic CPMG control sequence, various different types of DD control sequence were designed for different purposes\cite{Yang2010c}.
For example, the Uhrig's DD (UDD) sequence with nonuniformly-spaced pulses were proposed\cite{Uhrig2007} and experimentally\cite{Biercuk2009,Du2009} shown to be the optimal sequence to protect the short-time center spin coherence.
Also, various concatenated or nested schemes based on CPMG and UDD are developed for different purposes\cite{Yang2010c}, such as protecting spin coherence of multi-qubits\cite{Sun2010}. 
In this paper, aiming at identifying single nuclear spins and improving the resolution of nuclear spin noise spectroscopy, 
we generalize the standard CPMG sequence and design a new control pulse sequence.
With the designed sequence, more nuclear spins in the spin bath can be resolved in comparison to the standard CPMG sequence.

In order to demonstrate the application of our designed DD sequence, we focus on the identification of single nuclear spins around NV center electron spins in diamond,
which is a promising solid-state system in quantum information processing and nano-scale magnetometry\cite{Wrachtrup2006}.
We show that solely increasing the CPMG control pulse number does not help resolve more nuclear spins.
Instead, in order to achieve high resolution, in the designed pulse sequence, we replace the CPMG repetition unit (a single $\pi$-pulse) by a group of $\pi$-pulses.
The pulse timing structure within each repetition unit provides additional degrees of freedom to tailor the corresponding noise filter function in the frequency domain.
Thus, according to the features of the detected spin noise, the filter function can be fine-tuned to resolve more nuclear spins.
The principle of the DD sequence design can be used in the general noise spectroscopy in other systems such as trapped ions\cite{Biercuk2009} and superconducting qubits\cite{Bylander2011}.

This paper is organized as follows. Section \ref{SectII} describes the NV center spin coherence under DD control sequence and analyzes the semi-classical nuclear spin noise spectrum. 
Section \ref{SectIV} demonstrates the improved resolution of the designed DD sequence.
We conclude the paper in Section \ref{SectV}.


\section{NV center electron spin coherence under DD control}
\label{SectII}

\begin{figure}[tb]
  \includegraphics[width= 0.9\columnwidth]{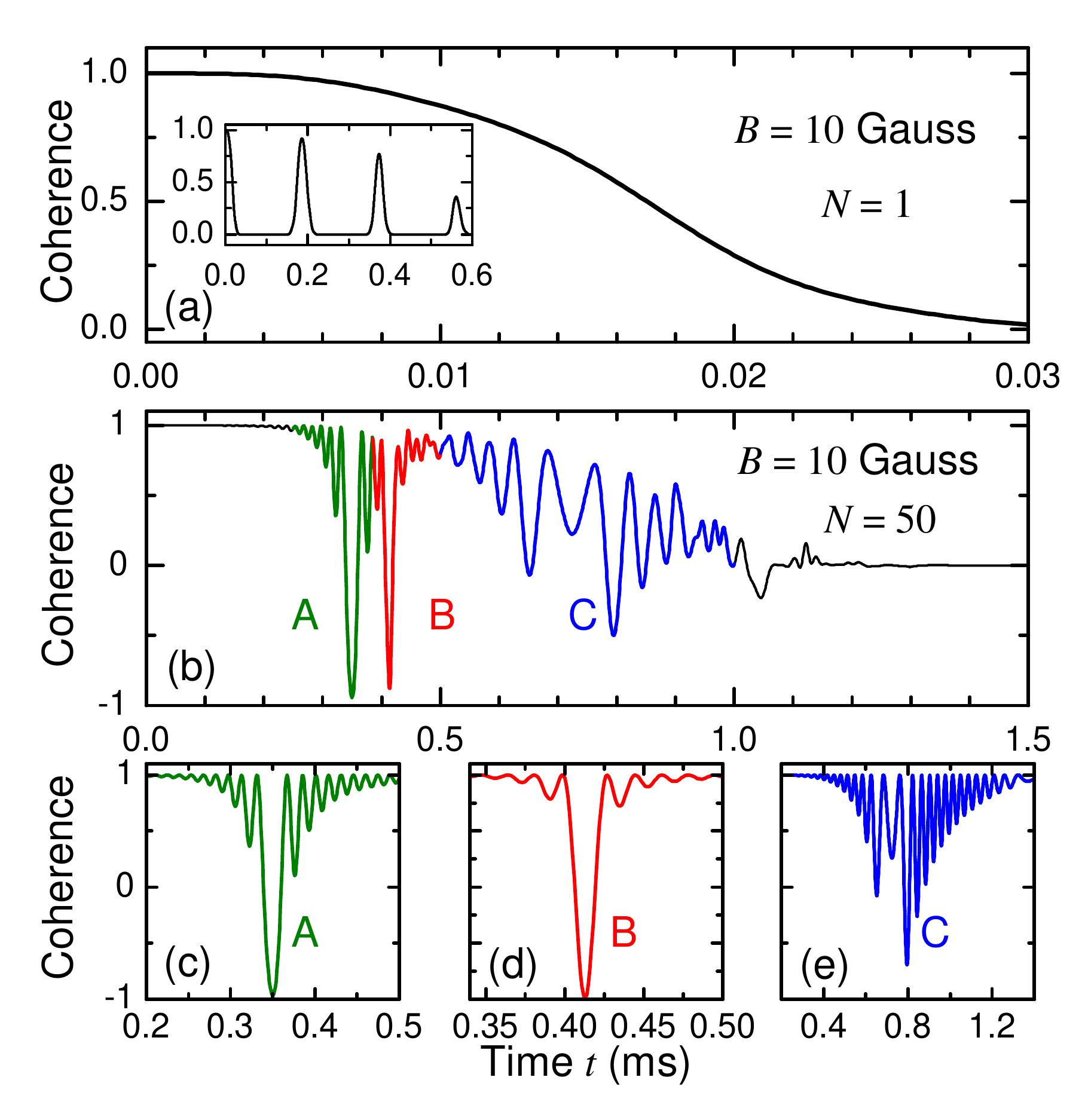}
  \caption{
  (a) The spin coherence under Hahn echo control in a $B=10$~Gauss magnetic field. 
  The inset shows the coherence collapse and revival in a longer time scale, and the main panel zooms in the first collapse.
  (b) The spin coherence under 50-pulse CPMG control. Only the first collapse is shown.
  Oscillations in green, red, and blue (denoted by A, B, and C in turn) are caused by three individual $^{13}$C spins with hyperfine coupling strengths $\sim 100$~kHz.  
  Their contributions to the coherence are singled out in (c)-(e) with corresponding colors.
  }
\label{FIG:fig1_CPMG}
\end{figure}

We consider the NV center system as an example to demonstrate the principle of DD design for high-resolution spin noise spectroscopy and nuclear spin identification. 
The NV center electron spin in a $^{13}$C nuclear spin bath is described by\cite{Maze2008a, Zhao2012}
\begin{equation}
H=H_{\text{NV}}+H_{\text{int}}+H_{\text{nuc}},
\label{Eq:full_hami}
\end{equation}
where the NV center Hamiltonian $H_{\text{NV}}$ is
\begin{equation}
H_{\text{NV}}=\Delta S_{z}^{2}-\gamma_{\text{e}}\mathbf{B}\cdot\mathbf{S}\equiv\sum_{\alpha=1}^{3}\omega_{\alpha}\vert\alpha\rangle\langle\alpha\vert.
\end{equation}
Here, $\mathbf{S}$ is the spin-1 operator of the electron spin, $\Delta=2\pi\times 2.87$~GHz is the zero field splitting, $\mathbf{B}$ is the applied magnetic field, 
and $\gamma_{\text{e}}=-1.76\times 10^{11} \text{rad } \text{s}^{-1} \text{ T}^{-1}$ is the electron spin gyromagnetic ratio. 
The $z$ direction is chosen along the N-V axis (the [111] direction).
In the eigen-representation, $H_{\text{NV}}$ is diagonalized with the eigen-frequencies $\omega_{\alpha}$ and the corresponding eigenstates $\vert\alpha\rangle$.

The Hamiltonian of $^{13}$C nuclear spins bath is 
\begin{equation}
H_{\text{nuc}}=-\gamma_{\text{C}}\mathbf{B}\cdot\sum_{j}\mathbf{I}_j+H_{\text{dd}},
\end{equation}
where $\mathbf{I}_j$ is the $j$th nuclear spin, $\gamma_{\text{C}}=6.73\times 10^{7} \text{rad } \text{s}^{-1} \text{ T}^{-1}$ is the gyromagetic ratio of $^{13}$C, 
and $H_{\text{dd}}$ describes the dipole-dipole interaction between $^{13}$C nuclear spins.
The electron spin couples to the $^{13}$C nuclear spins through the hyperfine interaction 
\begin{equation}
H_{\text{int}}=\mathbf{S}\cdot\sum_{j}\mathcal{A}_j\cdot\mathbf{I}_j=\sum_{\alpha, j}\vert \alpha\rangle\langle\alpha\vert \otimes \mathbf{A}_j^{(\alpha)}\cdot \mathbf{I}_j, 
\label{Eq:HF_int}
\end{equation}
where $\mathcal{A}_j$ is the hyperfine interaction tensor of the $j$th $^{13}$C, 
and $\mathbf{A}_j^{(\alpha)}=\langle \alpha \vert \mathbf{S}\vert \alpha\rangle\cdot\mathcal{A}_j$ is the hyperfine field felt by the $j$th nuclear spin for the electron being in state $\vert\alpha\rangle$. 
In the second equation of Eq.~\ref{Eq:HF_int}, we have neglected the electron spin flipping terms between different eigenstates $\vert\alpha\rangle$ and $\vert\beta\rangle$, 
since the electron spin is hardly flipped by the nuclear spins due to the large energy mismatch ($\sim$~GHz) compared with the typical hyperfine coupling strength ($<$~MHz).

In the NV center system described by the Hamiltonian (\ref{Eq:full_hami}), single nuclear spin detection has been realized in different magnetic field regimes\cite{Kolkowitz2012a,Taminiau2012a,Zhao2012}. 
%
In this paper, we will focus on how to improve the resolution in the weak field regime ($ B \sim 10\text{~Gauss}$, in diamond samples with natural abundance $^{13}$C isotope).
In this regime, single nuclear spin precession induces coherence collapse and revival effect in the Hahn echo.
The dipolar coupling $H_{\text{dd}}$ between nuclear spins causes the envelope decay of the revival peaks [see Fig~\ref{FIG:fig1_CPMG}(a)]\cite{Maze2008a, Zhao2012}.
Within the first collapse [for $t<30$~$\mu\text{s}$ in Fig.~\ref{FIG:fig1_CPMG}(a)], the dipolar coupling can be neglected and the nuclear spins precess independently.
In this case, the electron coherence between two eigenstates $\vert\alpha \rangle$ and $\vert\beta \rangle$ of $H_{\text{NV}}$ is expressed as\cite{Maze2008a, Zhao2012}
\begin{eqnarray}
&&L_{\alpha\beta}(t)=\prod_{j} L_{j,\alpha\beta}(t)\notag\\
&=& \prod_j \text{Tr}\left[\cdots e^{-i H_j^{(\alpha)}\tau_2} e^{-i H_j^{(\beta)}\tau_1}e^{i H_j^{(\alpha)}\tau_1}e^{i H_j^{(\beta)}\tau_2}\cdots\right],
\end{eqnarray}
where $L_{j,\alpha\beta}(t)$ is the contribution of the $j$th nuclear spin to the decoherence, and 
\begin{equation}
H_j^{(\alpha/ \beta)}=-\gamma_{\text{C}} \mathbf{h}_j^{(\alpha/ \beta)}\cdot\mathbf{I}_j
\end{equation}
is the conditional Hamiltonian of the $j$th nuclear spin in effective field 
$\mathbf{h}_j^{(\alpha/ \beta)} = \mathbf{B}-\mathbf{A}_j^{(\alpha/ \beta)}/\gamma_{\text{C}}$.

Fingerprint oscillations within the first coherence collapse emerge when increasing the CPMG control pulse number\cite{Kolkowitz2012a}.
Figure \ref{FIG:fig1_CPMG}(b) shows the calculated NV center electron spin coherence, in a randomly generated nuclear spin bath configuration, under 50-pulse CPMG control sequences
in a magnetic field $B=10$~Gauss along $z$ direction.
Under CPMG control, the first collapse time increases linearly as increasing the pulse number $N$\cite{Zhao2012}. 
Meanwhile, in contrast to the smooth decay within $\sim 30$~$\mu\text{s}$, strong characteristic oscillations appear on the decay profile [see the oscillations in green, red, and blue in Fig.~\ref{FIG:fig1_CPMG}(b)].
These oscillations are caused by three $^{13}$C nuclear spins close to the NV center, whose contributions are singled out in Figs.~\ref{FIG:fig1_CPMG}(c)-(e). The oscillation features (e.g., the positions and the depths of the coherence dips) contain the information of the hyperfine coupling, 
and have been used to resolve the single nuclear spins\cite{Kolkowitz2012a}.

\begin{figure}[tb]
  \includegraphics[width= \columnwidth]{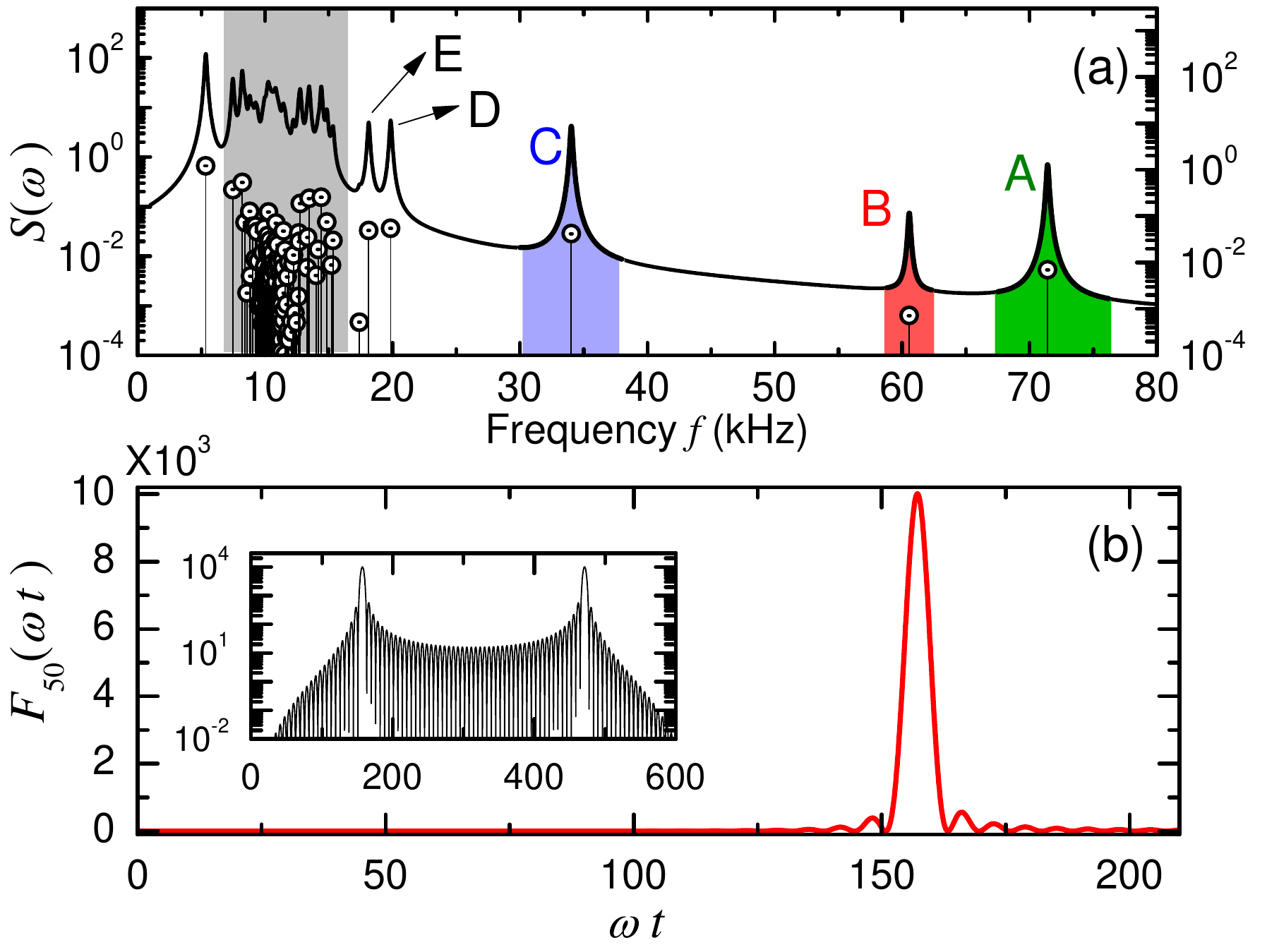}
  \caption{(Color online) (a) The noise spectrum of a nuclear spin bath coupled to an NV center spin. 
  Circle symbols represent the noise frequency and amplitude pairs  $(\nu_j, w_j)$ in Eq.~\ref{Eq:Spectrum}.
  The solid curve is obtained by replacing the $\delta$-function in Eq.~\ref{Eq:Spectrum} by a Lorentzian line-shape with a phenomenological broadening $\sim 0.1$~kHz (corresponding to the electron spin relaxation $T_1\gtrsim 1$~ms).
  A large number of $^{13}$C spins contribute to the quasi-continuous band (the grey shadow region) 
  around the $^{13}$C Larmor frequecy $f_{L}=10.7$~kHz at $B=10$~Gauss.
  Several $^{13}$C spins close to the NV center contributes to the discrete peaks.
  Three peaks, denoted by A (green), B (red), and C (blue) in turn, with  frequency higher than $f_{L}$ correspond to the oscillations shown in Fig.~\ref{FIG:fig1_CPMG}(c)-(e).
  Discrete peaks D and E close to the noise band are not resolved by the standard CPMG sequence by by the designed DD sequence (see Fig.~\ref{FIG:fig3_PackPulse}).
  (b) The filter function of $F_{50}(\omega t)$ 50-pulse CPMG control sequence. The inset shows a global view in logarithm scale.
  }
\label{FIG:fig2_spectrum}
\end{figure}

The underlying physics of the coherent oscillations can be understood in semi-classical picture with nuclear spin noise spectrum $S(\omega)$.
To calculate the noise spectrum, we define the averaged bath Hamiltonian $H_{\text{bath}}^{(j)}=\frac{1}{2}\left(H_j^{(\alpha)}+H_j^{(\beta)}\right)$ and the noise operator
$\hat{b}_j=H_j^{(\alpha)}-H_j^{(\beta)}$ of the $j$th $^{13}$C bath spin. 
The two eigen-frequencies and the corresponding eigenstates of $H_{\text{bath}}^{(j)}$ are denoted by $\nu_j^{(1,2)}$ and $\vert \psi_j^{(1,2)}\rangle$, respectively.
In general, the bath Hamiltonian $H_{\text{bath}}^{(j)}$ and the noise $\hat{b}_j$ are not commutative. The noise operator $\hat{b}_j$  will induce transition between the eigenstates $\vert \psi_j^{(1,2)}\rangle$.
Thus, the noise spectrum is calculated by
\begin{eqnarray}
S(\omega) = \sum_j w_j \delta\left(\omega-\omega_j\right),
\label{Eq:Spectrum}
\end{eqnarray}
where $w_j=\vert\langle \psi_{j}^{(1)}\vert \hat{b}_j\vert\psi_{j}^{(2)}\rangle\vert^2$ is the noise amplitude, and $\omega_j=\nu_j^{(2)}-\nu_j^{(1)}$ is the noise frequency.
Figure~\ref{FIG:fig2_spectrum}(a) shows the calculated nuclear spin noise spectrum of the same spin bath as that used in Fig.~(\ref{FIG:fig1_CPMG}).
The noise frequency is determined by both of the {\it intrinsic} Lammor frequency of $^{13}$C nuclear spins in the given magnetic field $\mathbf{B}$, 
and the {\it back-action} arising from the hyperfine coupling of each nuclear spins to the center electron spin\cite{Zhao2011b,Huang2011,Kolkowitz2012a}.
Particularly, for the center electron spin being prepared in the superposition state of $\vert 0\rangle$ and $\vert +1\rangle$, 
the $j$th $^{13}$C nuclear spin contributes noise with frequency $\omega_{j}=\vert-\gamma_{\text{C}}\mathbf{B}+\mathbf{A}_j/2\vert$.
The $^{13}$C nuclear spins far from the NV center electron spin have hyperfine coupling strengths much smaller than the applied magnetic field, 
i.e., $\vert \mathbf{A}_j\vert \ll \vert \gamma_{\text{C}}\mathbf{B}\vert$.
They produce nuclear spin noise with approximately the same frequencies, and form a quasi-continuous noise band around the Larmor frequency (see the grey shadow region in Fig~\ref{FIG:fig2_spectrum}).
The few $^{13}$C nuclear spins close to the NV center (with distance of several $\AA$) have hyperfine coupling strengths (in the order of $\sim 100$~kHz) comparable or larger than the applied magnetic field strength (but not strong enough to cause ESR spectrum splitting). 
Thus, they produce nuclear spin noise with frequencies significantly different from the Larmor frequency, and form discrete spectral lines.
These nuclear spins can be resolved by measuring the NV center electron spin coherence under DD control.

In the semi-classical picture, the coherence of a two-level system in a time-dependent noise field is approximately expressed as\cite{Cywinski2008}
\begin{equation}
L(t)\approx\exp \left[-\frac{1}{2}\int_{0}^{\infty}\frac{d\omega}{\pi} \frac{S(\omega)}{\omega^2}F_N(\omega t) \right],
\label{Eq:GaussianCoherence}
\end{equation}
where the noise filter function $F_N(\omega t)$ is the Fourier transform of the modulation function associated with the DD control scheme.
Figure~\ref{FIG:fig2_spectrum}(b) shows the filter function of the 50-pulse CPMG control. 
The periodic applied pulses in the time domain give rise to sharp peaks at frequencies $\omega_{N,k}= (2k-1)N\pi/t$ (for $k=1,2,\dots$).
As increasing the total evolution time $t$, the filter function peaks sweep from the high frequency side. 
Whenever the filter function peaks overlap with a discrete peak in the noise spectrum, there appears a coherence dip according to Eq.~(\ref{Eq:GaussianCoherence}).
When the first peak of the filter function touches the quasi-continuous noise band, the electron spin coherence collapses.


\section{Identification of single nuclear spins}
\label{SectIV}

\begin{figure*}[tb]
  \includegraphics[width= 1.7\columnwidth]{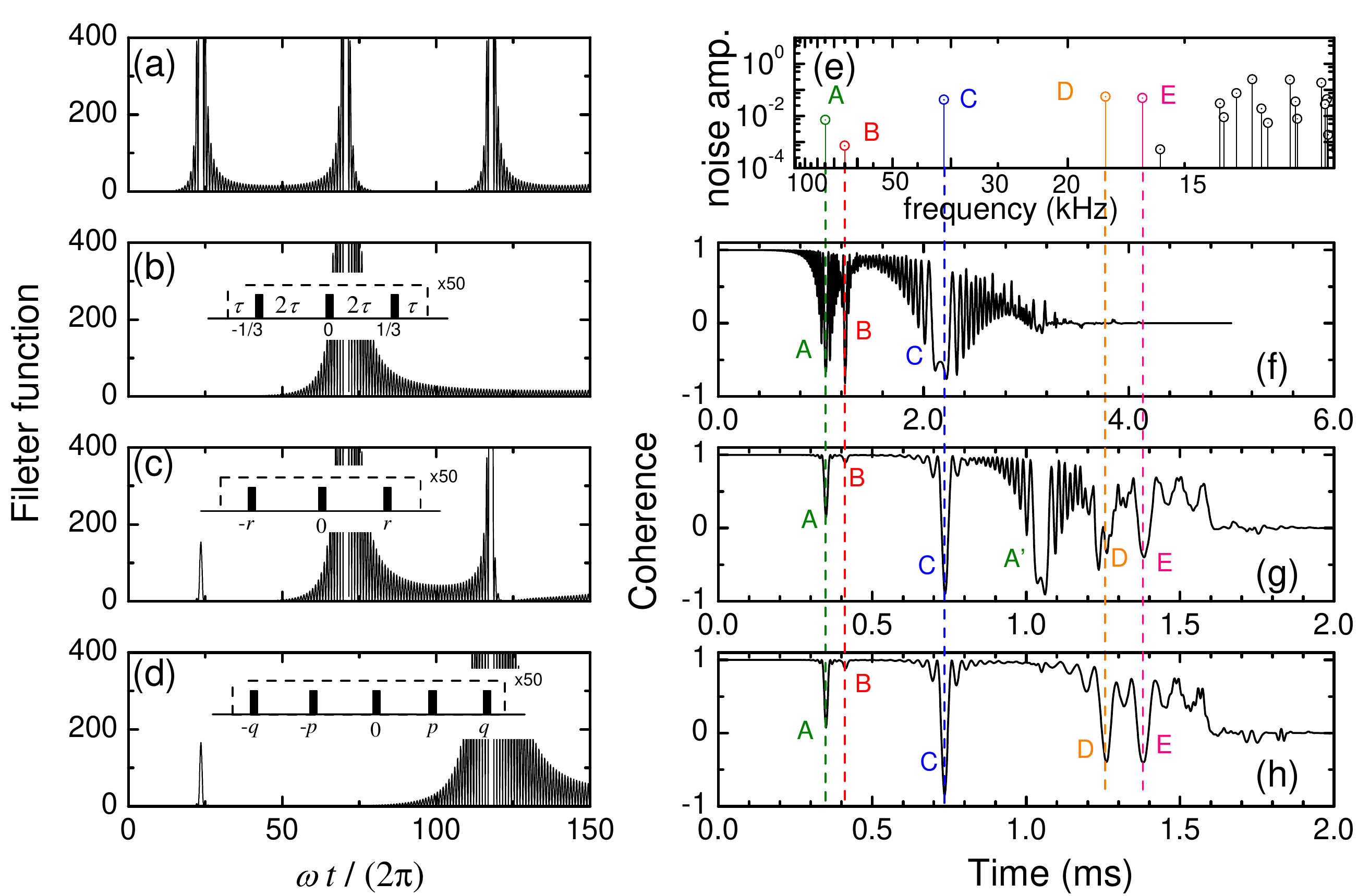}
  \caption{(Color online) 
  (a) The filter function $F_{50}(\omega t)$ of 50-pulse CPMG control.
  (b) The filter function $F_{150}(\omega t)$ of 150-pulse CPMG control. 
  The 150-pulse sequence can be regarded as 50 repetitions of 3-pulse unit shown in the inset.
  (c) The filter function of a modified $50\times 3$-pulse sequence. The repetition unit is shown in the inset with parameters $r=0.31$.
  (d) The same as (c) but for a modified $50\times 5$-pulse sequence. The parameters are $p=0.209$ and $q = 0.384$.
  (e) The noise spectrum in reciprocal axis. 
  (f)-(h) The electron spin coherence under the DD control with sequencies shown in (b)-(d) in turn.
  }
\label{FIG:fig3_PackPulse}
\end{figure*}
The results presented in Fig.~\ref{FIG:fig1_CPMG} seem to imply that increasing the coherence time by DD control would benefit the single nuclear spin identification. 
To check this, we compare the decoherence behavior under 50-pulse CPMG control with that under 150-pulse control, 
which has 3 times longer coherence time than the former (see Fig.~\ref{FIG:fig3_PackPulse}).
However, the nulcear spin resolution is not improved.
The same three nuclear spins as in the 50-pulse case are resolved from the coherence oscillation features.

Indeed, the unnecessarily strong peak in the filter function $F_N(\omega t)$ prevents the further improvement of the resolution.
Here, we should emphasize that Eq.~(\ref{Eq:GaussianCoherence}) is obtained by keeping the noise correlation only to the second order and, 
essentially, is a perturbation treatment valid only in the weak noise case.
When the filter function peak is too strong (i.e., the filtered noise $S(\omega)F_N(\omega t)/\omega^{2} \gg 1$, like in the 150-pulse CPMG case), 
the electron spin coherence will obviously deviate from Eq.~\ref{Eq:GaussianCoherence}.
The central dip is deformed and the side dips become strong. 
In this case, a single nuclear spin produces a broadened and highly oscillating structure in the electron spin coherence.
The interference of oscillating structures blurs the fingerprint features of individual nuclear spins to be resolved.


To solve this problem, we design pulse sequences simultaneously with high frequency selectivity and moderate peak strength of the filter function.
To this end, a group of $\pi$ pulses are used as the repetition unit, instead of a single one in CPMG sequence.
We consider an $N\times M$ sequence ($N$ repetitions of an $M$-pulse unit), whose pulses are applied at 
$t_{m,n}=[(2n-1)/2+r_m](t/N)$, where $n=1,2,\dots,N$ and $r_m\in [-\frac{1}{2},\frac{1}{2}]$ for $m=1,2,\dots,M$ is the relative position of the $m$th pulse within a repetition unit.

The principle of the designed pulse sequence can be understood through the analogy with the optical grating effect.
The periodic $N$ repetitions in time domain gives rise to the peak structure in the frequency domain described by
\begin{equation}
\tilde{F}_N(\omega t)=\frac{\sin^2 \frac{\omega t}{2}}{\cos^2\frac{\omega t}{2N}}.
\end{equation}
The structure {\it within} each repetition unit, which is local in time domain, imposes a slow modulation function $g_M(\omega t)$ of the peak structure in frequency domain.
The filter function is expressed as
\begin{equation}
F_{N\times M}(\omega t)=g_M(\omega t)\tilde{F}_N(\omega t),
\end{equation}
For the standard CPMG sequence ($M=1$), $g_{M=1}(\omega t)=16\sin^4[\omega t / (4N)]$.

Tuning the relative positions $r_m$ within the repetition unit modifies the strength of the filter function peaks.
Taking the symmetric 3-pulse repetition unit for example, with $-r_1=r_3\equiv r$ and $r_2=0$, we have
\begin{equation}
g_{M=3}(\omega t)= 16\left( \cos^2 \frac{\omega t}{4N} -\cos \frac{r\omega t}{N}\right)^2.
\end{equation}
As a special example, the 150-pulse CPMG sequence can be regarded as 50 repetitions of a 3-pulse unit with $r^{*} = 1/3$.
In this case, the pulses are indeed equally spaced, and one can check that the zero points of $g_{M=3}(\omega t)$ with $r^{*}=1/3$ 
precisely cancel the extra peaks of $F_{50}(\omega t)$ leaving only those peaks fulfil the peak condition of $F_{150}(\omega t)$ [i.e., $\omega t/(2\pi )=150\times(2k-1)/2$, see Fig.~\ref{FIG:fig3_PackPulse}(b)].

Slightly shifting the positions $r_m$ from the CPMG values (e.g., $r^{*}=1/3$ in the $M=3$ example above), we obtain a modified pulse sequence with both high frequency selectivity 
and moderate height of the filter function [see Figs.~\ref{FIG:fig3_PackPulse}(c)].
With this modified pulse sequence, as shown in Fig.~\ref{FIG:fig3_PackPulse}(g), two more nuclear spins (denoted by D and E) around the NV center are resolved.
As the frequencies of the noise produced by nuclear spins D and E are too close to the quasi-continuous noise band [around the Larmor frequency $\sim 10$~kHz for $B=10$~Gauss, see Figs.~\ref{FIG:fig2_spectrum}(a) and \ref{FIG:fig3_PackPulse}(e)], nuclear spins D and E cannot be resolved by the strong filter function peaks of the standard CPMG sequence.
While the much weaker and narrower filter function peak of the designed pulse sequence is able to resolve much finer structure close to the noise band edge.

More pulses in a repetition unit provides more degree of freedom to design the noise filter function. 
For example, the strong peak around $\omega t/(2\pi )\approx 75$ in Fig.~\ref{FIG:fig3_PackPulse}(c) causes a broad and strongly oscillating structure, denoted by $A^{\prime}$ in Fig.~\ref{FIG:fig3_PackPulse}(g).
Practically, this highly oscillating structure $A^{\prime}$ may interfere the identification of signals from other nuclear spins. 
Using a symmetric 5-pulse repetition unit shown in Fig.~\ref{FIG:fig3_PackPulse}(d), with relative pulse positions $-r_1=r_5=q$, $-r_2=r_4=p$ and $r_3=0$, one can eliminate the unwanted strong peak around $\omega t/(2\pi )\approx 75$, 
and get a much cleaner high-resolution fingerprint structure of five nuclear spins [see Fig.~\ref{FIG:fig3_PackPulse}(h)].

\section{Conclusion}
\label{SectV}
In this paper, we investigate resolving weakly coupled single nuclear spins from a nuclear spin bath.
A new type of DD control pulse sequence of central electron spin is designed. 
Applying the designed pulse sequence in NV center system, one can improve the resolution of the noise spectroscopy and, in comparison to the standard CPMG sequence, more nuclear spins around the NV center can be identified as resource for quantum information processing.
The principle of designing DD control pulse sequence is widely applicable in other systems, such as trapped ions and superconducting qubits, for high-resolution noise spectroscopy.

\section*{Acknowledgement}
This work is supported by Hong Kong RGC/CRF CUHK4/CRF/12G and CUHK Focused Investments Scheme. N.Z. acknowledges NKBRP (973 Program) 2014CB848700, and NSFC No. 11374032.
J.W. acknowledges funding from the EU via SIQS and SQUTEC as well as the DFG via FOR1493 and SFB/TR21 and BMBF via Q.COM.

\bibliographystyle{apsrev4-1}
%
\end{document}